\documentclass[final,5p,times,twocolumn]{elsarticle}

\usepackage{graphicx}

\usepackage{setspace}
\usepackage{enumerate}
\usepackage{times}
\usepackage{url}
\usepackage{graphicx}
 
\usepackage{amsmath}

\usepackage{float}
\usepackage{textcomp}

\usepackage{colortbl}
\usepackage{tabu}

\definecolor{darkblue}{rgb}{0,0,1}
\definecolor{darkgreen}{rgb}{0,0.6,0}
\definecolor{darkbrown}{rgb}{0.2,0.2,0.5}
\definecolor{darkblue}{rgb}{0,0,1}
\definecolor{darkmagenta}{rgb}{0.5,0,0.5}
\definecolor{lightblue}{rgb}{.7,0,1}
\definecolor{white}{rgb}{1,1,1}

\floatstyle{ruled}
\newfloat{algorithm}{thp}{lop}
\floatname{algorithm}{Algorithm}
\def\ds{\displaystyle}

\newcommand{\acrfont}{\fontfamily{qcr}\selectfont }

\def\ds{\displaystyle}

\begin{document}

\begin{frontmatter}

\title{An object oriented parallel finite element scheme for computations of PDEs: \\Design and implementation}


\author[label1]{Sashikumaar Ganesan\corref{cor1}}
\ead{sashi@cds.iisc.ac.in}
\author{Volker John$^{2,3}$}
\author{Gunar Matthies$^4$}
\author[label1]{Raviteja Meesala}
\author[label1]{Shamim Abdus}
\author{Ulrich Wilbrandt$^2$}
\address{$^1$Department of Computational and Data Sciences, Indian Institute of Science, Bangalore - 560012, India}
\address{$^2$Weierstrass Institute for Applied Analysis and Stochastics,  Mohrenstr. 39, 10117 Berlin, Germany}
\address{$^3$Department of Mathematics and Computer Science, Free University of Berlin, Arnimallee 6, 14195 Berlin, Germany}
\address{$^4$Department of Mathematics, Institute of Numerical Mathematics, Technical University Dresden, 01062 Dresden, Germany}

\cortext[cor1]{Corresponding author}

\begin{abstract}
Parallel finite element algorithms based on object-oriented concepts are presented. 
Moreover, the design and implementation of a data structure proposed are utilized in realizing a parallel geometric multigrid method. The ParFEMapper and the ParFECommunicator are the key components of the data structure in the proposed parallel scheme.  These classes are constructed based on the type of finite elements (continuous or nonconforming or discontinuous) used. The proposed solver is compared with the open source direct solvers, MUMPS and PasTiX. Further, the performance of the parallel multigrid solver is analyzed up to 1080 processors. The solver shows a very good speedup up to 960 processors and the problem size has to be increased in order to maintain the good speedup when the number of processors are increased further. As a result, the parallel solver is able to handle large scale problems on massively parallel supercomputers. The proposed parallel finite element algorithms and multigrid solver are implemented in our in-house package ParMooN.
\end{abstract}

\end{frontmatter}

\section{Introduction}
 Many physical phenomena and industrial processes are modeled by a set of partial differential equations (PDEs), and in many cases these PDEs are coupled
 and nonlinear in nature. Obtaining analytical solution of these  PDEs is very challenging   and impossible in most of the models. Therefore, the numerical  solution of    PDEs is of great interest in scientific and industrial applications. Advances in numerical methods for the solution of   PDEs facilitates to understand the physics of the problem   better, and to optimize the production in industries. Consequently, the computational complexity and   cost are also increased, and it necessitates efficient  numerical algorithms and implementations.  In several large scale applications, the use of supercomputer is inevitable. In recent years,   supercomputers are built with multicore processors, for e.g., the fastest supercomputer, as on June'16, Sunway TaihuLight consists of 40,960 processors with 256 processing cores each, that is, 10,649,600 CPU cores   in total. Moreover, CPU clusters are combined with Graphics Processing Unit (GPU) based accelerator clusters to gain performance and/or energy efficiency. For example, the supercomputer SahasraT at SERC, Indian Institute of Science, Bangalore, which is the fastest supercomputer in India, as on June'16, consists of 33,024 CPU cores and two accelerator clusters one with Nvidia GPU cards (44 nodes) and the other with Intel Xeon-Phi cards (48 nodes).
 In order to utilize the full potential of supercomputers and to achieve petascale and exascale computing in practical applications, the parallel algorithms need to be redesigned and re-implemented to support heterogeneous computing.
 


In general, the set of PDEs are discretized in space by the finite difference or finite volume or finite element method or one of its variants.
The finite dimensional discretization results in a large sparse system of  (mostly linear) algebraic equations.  In general, solving
the large sparse system accounts more than 90\% of the total computing time, and thus the scalability of the parallel implementations mainly depends on the scalability of the algebraic solvers used in the numerical scheme.  Apart from the other challenges associated with the parallel solution of sparse systems,   parallel computations require not only efficient parallel algorithms, but also highly scalable numerical methods. For instance, a stabilized numerical scheme with a local cell/matrix dependent stabilization parameter will be more efficient in parallel computations than a stabilized numerical scheme with a global mesh/matrix dependent stabilization parameter. Also, the choice of finite elements in finite element discretizations will influence the parallel efficiency. For example, the communication between the processors will be less when  non-conforming or discontinuous finite elements are used instead of continuous finite elements. 
Even though out-of-box solvers (e.g. CG, GMRES) work without any information about the underlying model problem and the numerical scheme, solvers
that aware the model and numerical scheme need to be developed in order to achieve a good performance in massively parallel supercomputers.


In general, the solvers that are used to solve a sparse algebraic system can be classified into two categories (i) Direct solvers and (ii) Iterative solvers. Some of the popular open source (academic) 
parallel direct solvers that support Message Passing Interface (MPI) are  MUMPS,  PaStiX,  PSPASES, PARDISO, SuperLU-DIST, WSMP.  Note that  the above
list is not complete, however, these are the commonly used solvers. Steps involved in a direct solver are the ordering of the linear system to reduce the fill-in,
the symbolic factorization, the numerical factorization and the solving step. Among all, the numerical factorization step is   computationally expensive. In direct solvers, 
the memory requirement increases due to  fill-in, when the problem size increases. This dependency on the fill-in could be observed when higher order finite elements are used or the dimension of the problem increases.
In a worst cases, where fill-in hinder the sparsity more or less entirely, the triangular solve alone could be $\mathcal{O}(n^2)$.
Therefore, the iterative solvers are preferred for very large systems. 

Unlike direct solvers,  only a very few open source iterative solvers are available, for example, HIPS, pARMS and Hypre. The iterative methods can further be classified into (a) stationary methods
(Jacobi, Gauss--Seidel, SOR, etc) and (b) instationary or Krylov subspace methods (CG, GMRES, BiCGSTAB, etc) see \cite{SAAD02} for more details.
To improve the convergence rate of the iterative solvers by reducing the condition number of the matrix, often the preconditioning technique is used. Popular preconditioner
are ILU, SOR, algebraic multigrid (AMG), geometric multigrid (GMG), etc. Among all, the multigrid method is very efficient, in particular for elliptic problems, and has 
$\mathcal{O}(n)$ complexity, where $n$  is the number of equations in the algebraic system \cite{chow2006survey,Joh02}.
Even though the multigrid method can be used as an iterative solver, often it is used as a preconditioner for the GMRES or other iterative methods. 
Construction of a geometric multigrid solver or a preconditioner for a parallel iterative solver is very challenging, as it requires communication between a hierarchy of 
distributed meshes. Also, the implementation of restriction and prolongation operators on a hierarchy of distributed meshes increases the complexity. Though many parallel solvers (mostly direct solvers) are available in public for large scale computing, most of these solvers do not support heterogeneous computing. 

In this paper we present a design and implementation of an object-oriented parallel finite element scheme that supports heterogeneous computing in addition to different types of finite elemnts.
The main objective of this work is to develop a parallel   solver  that is capable of  solving
large scale problems  so as to harness the massive computation capability of modern supercomputers.
To achieve this we reduce the communication overhead at every step and strengthen the 
algorithmic scalability at the same time. By algorithmic scalability we mean that the convergence rate of the solver does not degrade with the increase in the number of processor.  
The proposed implementation handles a hierarchy of finite element spaces defined on the hierarchy of distributed meshes. Further, a  parallel finite element
communicator class that automatically manages different finite elements (continuous, nonconforming and discontinuous) is implemented. Based on the  finite element
communicator, a parallel degree of freedom (DOF) class is implemented to handle the communication between the processors. Further, we   reduced the communication volume across processors considerably 
by implementing new data structures for mapping the interface nodes across processors. Moreover, the option of performing two or more smoothing iterations before
communicating with the neighboring processors is tested to optimize the ratio of communication and computations. Finally, the implemented parallel solver is compared with   MUMPS and PasTiX direct solvers. 

\section{FEM and iterative methods}
Principle steps involved in realizing FEM implementations are the assembling of the algebraic system and solving it.
In the cell based FEM approach, the system is assembled by looping over cells sequentially. This approach requires a numbering scheme for the degrees of freedom (unknown solution coefficients) in each cell locally and globally. The complexity increases when the parallel implementation is considered.
Consider a Gauss siedel iterative solver for a system $Ax=b$, the compute step
can be summarized as,
\begin{equation}\label{GS}
 x_{i}^{(k+1)} = b_i - \sum_{j=1}^{i-1}a_{ij}x_{j}^{k+1} - \sum_{j=i+1}^{n}a_{ij}x_{j}^{k},
\end{equation}
where $k$ is the index of the iteration step.
The sparsity in the system reduces the floating point operations  on the right hand side significantly, almost to a constant. In order to realize parallelism, one needs to distribute
this compute step across multiple processes. This involves redistribution  of the unknowns to balance the compute and communicate steps, without 
affecting the convergence of the algorithm.
The following sections address the design of data structures, in order to achieve parallelism for such numerical algorithms. 
\section{Object oriented Finite Element methods}
Object oriented approach for finite element methods have been favoured since the 1990's\cite{FORDE1990355}. 
The strength of the approach lies in the modularity achieved and the net decrease in the lengths of code.
The advantages of these techniques can be found in \cite{Mackerle2004325} \cite{doi:10.1108/eb023638} \cite{Dadvand2010}.
In general, such a code can broadly be  divided into four main parts.
\begin{itemize}
 \item Domain decomposition (meshing, mesh partition)
 \item Construction of finite element structures (DOFs, Matrix stencils)
 \item Assembling of system matrices
 \item Solving  system of equations
\end{itemize}
The following section is focused on describing   parallel data structures that are required for  implementations of the above steps.

\section{Parallel Data Structures}\label{refer1}
\subsection{Mesh Partitioning}
The implementation begins from importing the geometry (or) the domain of the problem. The domain can be discretized either internally or by using  external mesh generation packages such as Gmsh~\cite{geuzaine2009gmsh}. Distributing the mesh cells across processors  helps in achieving coarse grain parallelism. 
Several strategies can be thought out to partition the mesh across the processes.
Suppose the mesh is distributed more or less uniformly across all processors (or with respect to the number of nodes), we can achieve a good load balancing in computation. Another strategy could be to try and minimize 
the "interface" area that results due to partitioning. It affects the amount of communication that takes place across the processes. One of the most popular 
packages used for handling this task is METIS~\cite{karypis1995metis}. Each process is allocated a 'subdomain' (a collection of cells in the mesh) on which it performs the computation. Further refinement of the mesh can be performed parallely by each process
over their own corresponding subdomains.

\subsection{ParFEMapper - Parallel Finite Element Mapper}
The   degrees of freedoms (DOFs) of a three-dimensional (3D) finite element  might be defined on the vertices, edges, faces and/or interior of the mesh cells based on the types of finite elements (continuous or nonconforming or discontinuous) used to construct the  finite element space (FESpace).
ParFEMapper is a class containing the mapping information of DOFs on the subdomain interface, and it facilitates to communicate solutions on the interfaces of subdomains between processors.

\subsubsection{Cell nomenclature}  
\begin{figure}
\begin{center}
\includegraphics[width=0.4\textwidth]{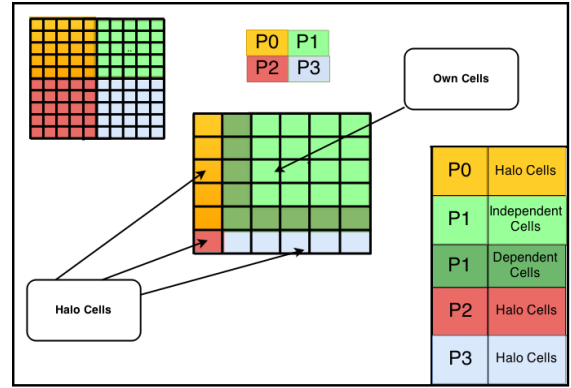}
\caption{Cell Nomenclature in the subdomain of the processor  P1.}
\label{CellNomenclature}
\end{center}
\end{figure}

METIS  partitions the mesh and assigns a processor number to each cell. Then this information can be broadcasted to all MPI processors. After that each MPI processor collects the set of cells with its own processor number (rank), and marks all these cells as {\acrfont Own Cells}. Further, the {\acrfont Own Cells} are divided into {\acrfont Dependent} and {\acrfont Independent Cells}, where the set of all own cells that are connected with the
neighboring MPI processors' cells are called {\acrfont Dependent Cells}. The remaining  own cells  are called {\acrfont Independent Cells}, which do not depend on the neighboring processors directly.  
Note that two cells from different MPI processors might be connected by a vertex or edge or face in 3D. Suppose a vertex or an edge or a face is shared by two 
or more cells from different MPI processors, we call it as a {\acrfont Subdomian Vertex, Subdomian Edge, Subdomian Face}, respectively. The collection of these subdomain vertices, edges and faces is called  {\acrfont Subdomian Interface}.
To calculate/update a DOF defined on  the subdomain interface, the corresponding MPI processor of this DOF  must contain all cells associated with this   
DOF,  and some of the associated cells must belong to neighboring processors. These additional associated cells are also   necessary to assemble a consistent distributed system. The associated cells that are in neighboring processors are  called {\acrfont Halo Cells} of the  corresponding MPI processor. 
Thus, a {\acrfont Halo Cell} on a MPI processor has a support for a DOF of the MPI processor, however the cell  is an own cell on its neighboring MPI processor. 
For example, conforming and nonconforming FESpaces will have different collection of Halo cells, since the DOFs of the nonconforming FESpace are not defined on vertices. 
Finally, the collection of {\acrfont Own Cells} and {\acrfont Halo Cells} together form a subdomain mesh for the respective MPI processor. 
Hence, the total number of cells on each MPI processor  is given by,
\begin{align*}
\acrfont 
\text{Total\_N\_Cells} ~= ~& \acrfont\text{N\_Own\_Cells} \\ 
                      & +~\acrfont\text{N\_Halo\_Cells},
\end{align*}
where the total number of {\acrfont Own Cells}  is given by, 
{\acrfont
\begin{align*}
\text{N\_Own\_Cells} ~= ~&\text{N\_Dependent\_Cells }  \\ & +~\text{N\_Independent\_Cells}.
\end{align*}}
Figure \ref{CellNomenclature} shows various types of cells in the subdomain of processor P1. Further, 
P0, P1, P2 and  P3 in Figure \ref{CellNomenclature} and \ref{DofNomenclature} denote different processors (ranks).

\subsubsection{DOF Nomenclature} 
Based on the choice (continuous or nonconforming or discontinuous)  of finite  elements, the  DOFs are defined on vertices, edges, 
faces  and interior of the cells. Some of these DOFs that are defined on the {\acrfont Subdomian Interface} will be shared by both {\acrfont Dependent Cells} and {\acrfont Halo Cells}.  
Such DOFs are called {\acrfont Interface DOFs}, that is, the set of all DOFs that are defined on the {\acrfont Subdomian Interface}. 
Since each  {\acrfont Interface DOF} belongs to more than one MPI processor,   one of the associated processors is given the responsibility of computing the solution at this DOF.
This computing MPI processor is called {\acrfont Master Processor} of this DOF. Further,  this interface DOF is called {\acrfont Master DOF} in the computing MPI processor,
whereas it is termed as a {\acrfont Slave DOF} on all other associated MPI processors.
In other words, the {\acrfont Interface DOF} is a {\acrfont Master DOF} on a MPI processor, if the processor takes the responsibility of 
computing the solution else it is a {\acrfont Slave DOF}. 

Next, the collection of DOFs that are defined on the {\acrfont Halo Cells} but not on the {\acrfont  Subdomian Interface} are called {\acrfont Halo DOFs}. The
{\acrfont Halo DOFs} are further divided into two categories - {\acrfont Halo1 DOFs} and {\acrfont Halo2 DOFs}.  Suppose a {\acrfont Halo DOF} is having a support (connection) 
with any of the {\acrfont Master DOFs}, then it is marked  as {\acrfont Halo1 DOF}, else marked  as {\acrfont Halo2 DOF}.

\begin{figure}
\begin{center}
\includegraphics[width=0.4\textwidth]{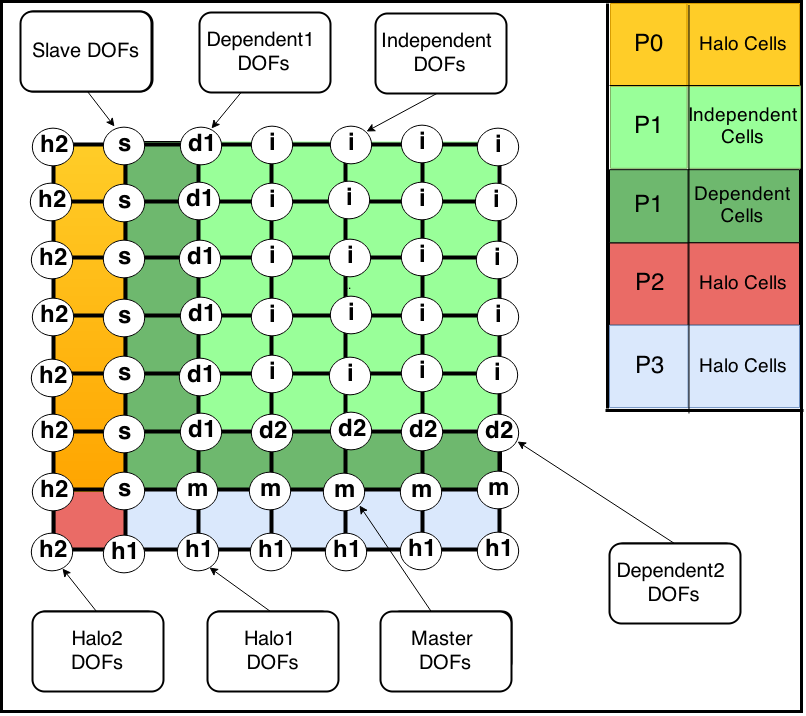}
\caption{DOF Nomenclature for $Q_1$ finite element on     MPI processor P1.}
\label{DofNomenclature}
\end{center}
\end{figure}

Furthermore, to enable hybrid (threads on each MPI processor) the DOFs are marked with different labels. The collection of DOFs that are defined on the {\acrfont Dependent Cells} 
but not on the {\acrfont  Subdomian Interface} are called {\acrfont Dependent DOFs}. 
The {\acrfont Dependent DOFs} are also further divided into two categories - {\acrfont Dependent1 DOFs} and {\acrfont Dependent2 DOFs}. Suppose a
{\acrfont Dependent DOF} is having a support (connection) with any of the {\acrfont Master DOFs}
then it is marked  as {\acrfont Dependent2 DOF},  else marked  as   {\acrfont Dependent1 DOF}.  The remaining  DOFs that are defined on the {\acrfont Independent Cells} of 
the subdomain are called {\acrfont Independent DOFs}.
Hence, on each MPI processor, we have
{\acrfont 
\begin{align*}
 \text{N\_DOFs} = &\text{ N\_Independent\_DOFs } \\
                  & ~+  \text{ N\_Dependent\_DOFs }  \\
                  & ~+ \text{ N\_Interface\_DOFs} + \text{ N\_Halo\_DOFs},
\end{align*}
}
where
{\acrfont 
\begin{align*}
\text{N\_Dependent\_DOFs} = & \text{ N\_Dependent1\_DOFs }   \\
                       &~ +\text{ N\_Dependent2\_DOFs} \\
\text{N\_Interface\_DOFs}  = & \text{ N\_Master\_DOFs }  \\ 
                            &~+\text{ N\_Slave\_DOFs}\\
\text{N\_Halo\_DOFs} = &\text{ N\_Halo1\_DOFs } \\
                      &~+ \text{ N\_Halo2\_DOFs}.
\end{align*}
}
Figure \ref{DofNomenclature} shows various types of DOFs in the subdomain of the MPI processor P1.
The benefits of this nomenclature can be fully realized in the ease of implementation of different compute strategies that could be adopted in a multigrid technique.

\subsubsection{Mapping of DOFs across processors}\label{mappingDOFs}
\begin{figure}
\begin{center}
\includegraphics[width=0.4\textwidth]{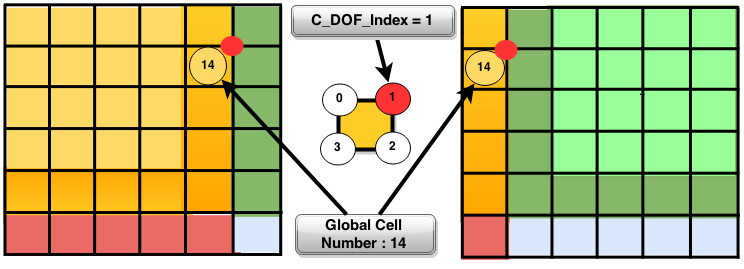}
\caption{Mapping of a slave DOF between two MPI processors.}
\label{dofmapping}
\end{center}
\end{figure}

 The DOFs are indexed (numbered) independently on each MPI processor while constructing the FESpace on their subdomain mesh. Thus, the indices of the  dependent, 
 interface and halo DOFs on a MPI processor will be different from the  indices assigned by the FESpaces of its neighboring MPI processors. 
 Therefore, the mapping for the dependent, interface and halo DOFs with their neighboring MPI processors needs to be constructed. 
 It is the main purpose of the  ParFEMapper class. To construct a map, we use the {\acrfont Global Cell Number} of the  cells. 
 Since the coarse mesh is same on all MPI processors before partitioning, the {\acrfont Global Cell Number} of a coarse cell is unique across all MPI processors.  
Let us first consider the mapping of Master-Slave DOFs. Since the subdomain contains all cells (including halo) associated with the interface DOFs, 
each interface DOF knows the ranks of all MPI processors associated with it. Using this information, 
the interface DOFs are first distributed (divided into Master and Slave) uniformly across all MPI processors to maintain the load balance. 
Next, we map the  {\acrfont Slave DOFs} in the MPI processor with their corresponding {\acrfont Master DOFs} in the neighboring processor. 
Consequently, all {\acrfont Master DOFs} will be mapped with the {\acrfont Slave DOFs}. To map a slave DOF, the following information associated with the Slave DOF are collected:
\begin{enumerate}
\item [a)]  {\acrfont{{{Global Cell Number}}}} -  the global cell number of the slave DOF
\item [b)]  {\acrfont{{{C\_DOF\_Index}}}} - the local cell index of the slave DOF in the respective global cell
\item [c)] {\acrfont{{{P\_DOF\_Index}}}} - the FESpace DOF index of the slave DOF in the MPI processor
\end{enumerate}   
These information are collected for all slave DOFs, and sent to the respective neighboring processors that consider  these interface DOFs as Master DOFs.
Note that more than one own cells might be associated with a slave DOF,
however it is enough to choose any one of the associated cells, and the corresponding local cell index in the chosen cell. 
Once this information is received, the master processor identifies its own cell for the received {\acrfont{{{Global Cell Number}}}}. 
Then, it maps their FESpace DOF index to the  received {\acrfont{{{P\_DOF\_Index}}}}  from the neighboring (slave) processor by matching their local cell index with
the received {\acrfont{{{C\_DOF\_Index}}}}. Finally, the master MPI processor sends this mapping to all slave MPI processors.
Figure \ref{dofmapping} shows the mapping of the red colored DOF between two MPI processors using this procedure.

We next consider the mapping of {\acrfont Dependent DOFs}  and {\acrfont Halo DOFs}. According to our DOF nomenclature, the {\acrfont Dependent DOFs} of a MPI processor 
are {\acrfont Halo DOFs} of their   neighboring processors. Therefore, it is enough to send the {\acrfont Dependent DOFs} and consequently the mapping for the {\acrfont Halo DOFs} is received. 
To map the {\acrfont Dependent DOFs}, we use the same procedure as described in the Master-Slave DOF mapping.  
While the global cells IDs used to map the interface DOF's across processes, the local IDs are also retained to perform Matrix vector operations efficiently on the local system matrices that are assembled.

\subsubsection{Halo1 DOFs and  Halo2 DOFs}
{\acrfont Halo DOFs} are divided into  {\acrfont Halo1 DOFs} and {\acrfont    Halo2 DOFs} based on its support with the interface DOF. During the solution process, 
each MPI processor computes/updates only the master, dependent and independent DOFs. Therefore, the updated values of only a few halo DOFs that are having support with 
the master DOFs are needed  during the iteration. These  Halo DOFs are marked as {\acrfont Halo1 DOFs}, and the remaining Halo DOFs marked as {\acrfont Halo2 DOFs}.
During the iteration, only the {\acrfont Halo1 DOFs} are communicated, whereas the {\acrfont Halo2 DOFs} are communicated only  before performing restriction and prolongation
operations in multigrid method. Further, the updated {\acrfont Halo2 DOFs} values are needed  when the solution is part of the matrix assembling in nonlinear or coupled  problems.

\begin{figure}
\begin{center}
\includegraphics[width=0.5\textwidth]{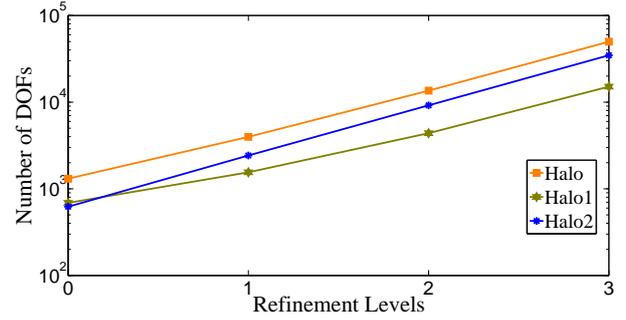}
\caption{Increase in the number of Halo, Halo1 and Halo2 DOFs on 6 MPI processors while increasing level of mesh refinement.}
\label{HaloDofs}
\end{center}
\end{figure}

Communicating only the values of {\acrfont Halo1 DOFs} rather than the values of all {\acrfont Halo DOFs} reduces the communication volume by a considerable amount. 
We can observe in Figure \ref{HaloDofs} that even for six MPI processors, the number of {\acrfont Halo2 DOFs} increases with an increase in the uniform refinement of
the mesh. Consequently, the difference between the number of {\acrfont Halo} and {\acrfont Halo1 DOFs} also increase. The difference becomes more significant with an increase
in the number of processors as the subdomain interface area increases.

\subsubsection{Dependent1 DOFs and Dependent2 DOFs}
The {\acrfont Dependent DOFs} and {\acrfont{Master DOFs}} of a MPI processor are  {\acrfont Halo DOFs} and  {\acrfont Slave DOFs}, respectively, on their neighboring processors. Further, we denote the set of all {\acrfont Dependent DOFs} that are  having support with the {\acrfont Slave DOFs} as {\acrfont Dependent1 DOFs}, 
and the remaining  {\acrfont Dependent DOFs} are marked as  {\acrfont Dependent2 DOFs}. 
By our convention, the {\acrfont Dependent1 DOFs} on a MPI processor are actually the {\acrfont Halo1 DOFs} for the neighbouring processor 
which is master of the slave DOFs connected to Dependent1 DOFs.
Hence, it is sufficient to send the updated values of {\acrfont Dependent1 DOFs} in order to update the values of {\acrfont Halo1 DOFs}.
Similarly, it is enough to send the values of {\acrfont Dependent2 DOFs} to update {\acrfont Halo2 DOFs}.

\subsubsection{DOF Reordering}
In order to access different sets of marked DOFs independently, the indices of FESpace DOFs are renumbered in the ParFEMapper. 
The {\acrfont Master DOFs} are numbered first followed by the
{\acrfont Independent, Dependent1, Dependent2, Slave, Halo1} and {\acrfont Halo2 DOFs} with the Dirichlet DOFs at the last. 
In addition, the {\acrfont Master, Dependent1, Dependent2} and {\acrfont Independent DOFs} are colored in hybrid computations, and further numbered color-wise.

\subsubsection{Local to Global Mapping for Direct Solvers}\label{mappingMUMPS}
Even though the DOFs are numbered locally on each MPI processor and the matrices are assembled in a distributed way, 
the parallel direct solvers such as MUMPS need the global row and global column indices of the entries in the distributed matrix.
To assign a global number for each DOF, each MPI processor calculates the number of  {\acrfont Own DOFs} as 
\begin{align*}
 \acrfont\text{N\_Own\_DOFs} = &\acrfont\text{ N\_Independent\_DOFs } \\
                  & ~+  \acrfont\text{ N\_Dependent\_DOFs }  \\
                  & ~+ \acrfont\text{ N\_Master\_DOFs}.
\end{align*}
Each MPI processor broadcasts its   {\acrfont N\_Own\_DOFs} and  creates an array  {\acrfont N\_Proc\_OwnDOFs} 
on all processors. For example, {\acrfont N\_Proc\_OwnDOFs[k]} has the {\acrfont N\_Own\_DOFs} of k$^{th}$ MPI processor. 
Using this array, the global DOF numbering of {\acrfont Own DOFs}  in the MPI processor {\acrfont\text{``p''}} is assigned  as 
\begin{align*}
{\acrfont\text{Global\_DOF[i]} = \ds\sum_{\text{k=1}}^{\text{p-1}} \text{ N\_Proc\_OwnDOF[k]  + i}},
\end{align*}
 for {\acrfont\text{i = 1,\ldots,\text{N\_Proc\_OwnDOF[p]}}}.
The halo  and slave DOFs  of the MPI processor  {\acrfont\text{``p''}}  receive their unique global DOF number from  their neighboring 
processors. 

\subsection{ParFECommunicator - Parallel Finite Element Communicator}
The ParFECommunicator class consists of various parallel communication methods that are implemented  on the basis of the mapping defined in ParFEMapper. These routines are used to communicate
information across processors while using multigrid and direct solvers. Communication using this class can be performed 
at various levels of meshes. These routines can also be used to communicate {\acrfont Master DOFs} and {\acrfont Slave DOFs} or/and {\acrfont Halo1 DOFs} 
or/and {\acrfont Halo2 DOFs}. 
Finally, we conclude this section with a note that ParFEMapper and ParFECommunicator depend on the used finite element spaces. 
Continuous, non-conforming and discontinuous finite elements will have different instances (objects)  of these classes on the same mesh. Further, these  objects  need  to be  generated only once at the beginning of parallel computations, and it is highly scalable. 

\section{Parallel Multigrid Solver}
In this section, we discuss the utilization of parallel data structures discussed above in realizing a parallel multigrid solver.
This is done in several steps. 
On a hierarchy of meshes in geometric multigrid method, the objects of the ParFEMapper and ParFECommunicator classes  need to be constructed on all mesh levels. 

\subsection{Construction of Hierarchy of Meshes in Parallel}
The subdomain mesh ({\acrfont Own Cells} and {\acrfont  Halo Cells}) in every MPI processor is uniformly refined till the finest level is achieved.  
The refinement of {\acrfont Halo Cells} at every mesh level generates new  cells (children), of which, some of them will not have any connectivity to the Dependent cells of the MPI processor. Therefore, the unwanted new children cells are removed from the new subdomain collection of cells. After that, a new FESpace is constructed on this new subdomain, 
and that is used to construct new  objects of the  ParFEMapper and ParFECommunicator classes. This process is repeated until the finest multigrid level.

\subsection{Global Cell Number in Hierarchy of Meshes}
Since the global cell number is used to construct a map in the ParFEMapper class, a global cell number has to be assigned for the newly generated children cells in the refinement. 
The {\acrfont Global Cell Number} across all processors will be unique 
only at the coarsest level. After partitioning the coarsest mesh, the refinement of the mesh is local to every processor.
We assume that the refinement is uniform on all MPI processors. Using this assumption and the global number of the parent cell, the global cell number of the $i^{th}$     child cell at $\l^{th}$ level is assigned as
\begin{equation*}
\acrfont\text{GCN$_{(l)}[i]$} =   \acrfont\text{NC}  \times \text{PGCN$_{(l)}[i]$} + \text{CI}, 
\end{equation*}
where GCN$_{(l)}[i]$  is the {\acrfont Global Cell Number} of the $i^{th}$  cell at level $l$, NC is the number of newly generated children of the parent cell,  
PGCN$_{(l)}[i]$ is the {\acrfont Global Cell Number} of the parent of the cell and CI is the local index of the child cell in the parent cell.
The above procedure guarantees a unique {\acrfont Global Cell Number} for all cells  when the mesh refinement is uniform on all MPI processors.  

\subsection{Parallel Multigrid Cycle}
Different multigrid cycles   are implemented in our in-house package ParMooN \cite{ParMooN}. Let us consider the V-cycle.
Each processor constructs an instance (object) of the multigrid solver class  after having generated a hierarchy of meshes. 
After the assembling of the system matrix, the multigrid solver starts to perform a few iterations on the finest mesh to smooth out the high frequency error components. 
After every iteration of the pre-selected iterative solver the updated values are communicated using the ParFECommunicator.  
This step is known as the pre-smoothing step. The residual of the fine system is then restricted to the coarser level. 
The residual equation is solved   on the coarser mesh. 
The  smoothing and restriction steps are repeated until the coarsest level is reached. 
The residual equation is solved exactly (or up to a predefined level of accuracy) on the coarsest level to get the update/correction. 
After that the update/correction is prolongated to the next finer level and added to the solution. 
Few more iterations using the predefined smoothers like Gauss-Seidel are run on the finer mesh by considering the new improved value as an initial guess.
This step is known as post-smoothing. The prolongation and post-smoothing operations are performed on each level till the finest mesh is reached.
Note that only the updated values of the {\acrfont Master DOFs} and {\acrfont Halo1 DOFs} are communicated at each iterative step of the smoothing operation. Further,  
the {\acrfont Halo2 DOFs} are   communicated before every restriction and prolongation operations.
The same algorithm can be used to run different multigrid  cycles.
\begin{algorithm}
\caption{Parallel Multigrid Solver : (V$  -$Cycle) }
\begin{enumerate} 
\item Repeat till coarsest level is reached
\begin{enumerate} 
\item Pre$-$Smoothing : Reduce high frequency errors by performing few iterations of Jacobi or Gauss--Seidel method. 
      At every iterative step of Jacobi or Gauss--Seidel communicate the defect on {\acrfont Master} and {\acrfont Halo1 DOFs} across processors using ParFECommunicator.
\item Communicate the {\acrfont Halo2 DOFs} after the pre-smoothing step.
\item Restriction : restrict the residual values from finer level to coarser level.
\end{enumerate}
\item Solve the residual equation exactly at the coarsest level either by using a direct solver or by using an iterative solver until convergence to solution is achieved.
\item Repeat till finest level is reached.
\begin{enumerate} 
\item Prolongation : prolongate the solution of the residual equation from the coarser level to the finer level and add it to the previous approximate solution. 
\item Post$  -$Smoothing : Reduce high frequency errors by performing few iterations of Jacobi or Gauss--Seidel method with the new improved initial guess. 
                           At every iterative step of Jacobi or Gauss--Seidel communicate the defect on {\acrfont Master} and {\acrfont Halo1 DOFs} across processors using ParFECommunicator.
\item Communicate the {\acrfont Halo2 DOFs} after the post-smoothing step.
\end{enumerate}
\end{enumerate}
\end{algorithm}
\subsection{Restriction and prolongation operators}
The restriction and prolongation operators determine the efficiency of the multigrid implementation. We use  a general  transfer operators proposed in \cite{Schieweck00ageneral}  for arbitrary  finite element spaces. In the case of parallel implementation, one needs to 
apply these operators with the help of the ParFECommmunicators to handle  interface and halo DOF's. The implementation utilizes the knowledge of own cells and master DOFs apart from the ParFECommmunicator discussed above.
Further, an additional restriction operation is required to assemble the system matrices at all levels  when multigrid methods are adopted for non-linear problems, such as   Navier-Stokes. It needs to be performed
whenever the solution is updated iteratively.

\subsection{Complexity}
\subsubsection{Computational complexity}
The steps involved in the iterative technique adopted is to perform a fixed point iteration
followed by a multigrid V or W cycle. Within a multigrid cycle, iterative sweeps are performed at 
each of the levels considered. Assume 'v' number of multigrid cycles are performed 
within a fixed point iteration. Let 'PRS' be the number of pre-smoothing steps performed on each level, before
performing a restrict operation and let 'POS' be the number of post-smoothing steps performed on each level, after
prolongate operation. Generally these two are chosen to be equal. Additionally one could perform multiple local sweeps 'L', before performing a 
communication update, in the case of a parallel implementation.
The complexity of a smoothing step, as in the solution step \eqref{GS}, is $\mathcal{O}$(N). Similarly, the restriction and prolongation operations are $\mathcal{O}$(N).
Hence, the total complexity in a multigrid sweep per level would be : v*(2*PRS)*L*$\mathcal{O}$($N_l$),
where $N_l$ is the total number of DOFs in a given level. In the case of 3D problems,
coarser levels will have $\approx$ $N_F/8$ DOFs, where $N_F$ is the number of DOFs on a finer mesh. 
This indicates that the total DOFs across all levels is bounded
by $\mathcal{O}$($N_F$)
Hence the total complexity can be approximated to be : v*(2*PRS)*L*$\mathcal{O}$($N$).
The parameters v, PRS, L can be chosen appropriately to affect the rate of convergence.

\subsubsection{Communication Vs Computation}
For the sake of analysis of the implementation,   consider a   cubic domain.
The partition is assumed to be uniform, i.e. each process obtains a sub-cube of same volume.
Consider a cube of side length A as our physical domain. The toal number of nodes N (number of nodes) $\approx$ $A^3$. 
Assume the domain is partitioned across K processes. This gives a (sub)cube of volume ($A^3$/K) and a  side length of  (A/$K^{1/3}$) $\approx$ $(N/K)^{1/3}$ for the subcube.
Since the interface determines the communication, we consider the faces of the cube which constitute the interface.
Surface area of (sub)cube $\approx$ c*$(N/K)^{2/3}$. This approximation holds for every sub-cube (having varying number of faces as interfaces).
Every cube now shares a boundary (edges or faces or corner vertices) with at most 26 neighbouring sub-cubes.
Since the interaction of the sub-cube with these neighbours is bounded by a constant, and the information is only required locally we can assume that the communication complexity is directly proportional to the calculated area of the interface $\approx$ $(N/K)^{2/3}$. 

Considering the ratio of computation to communication, we have: Computation $\approx$ c*(N/K) and Communication $\approx$ $(N/K)^{2/3}$. The ratio that we obtain is $\approx$  $\mathcal{O}$($(N/K)^{1/3}$).
Consider an embarassingly parallel program, with  communication $\mathcal{O}$(1), the ratio is $\approx$ $\mathcal{O}$(N/K). This implies that the algorithm becomes I/O bound at a quicker rate than an embarrassingly parallel program. 
However, the problem can be hard to scale linearly, i.e. with a ‘c1’ fold increase in size of the problem (through a higher level of refinement), we can scale the problem with ‘c1’ times more processes.
Also, the other way to achieve this is if we can increase compute by a factor of ‘c2’. Then again we can scale it with ‘c2’ times more processes.

\section{Numerical Results}

\subsection{ParMooN}
The above discussed parallel data structures and the parallel multigrid solver are implemented in our in-house package ParMooN \cite{ParMooN}.
It is built on MooNMD (Mathematics and object oriented Numerics in MagDeburg)   \cite{john2004moonmd}. These packages are built using Object Oriented C++. In addition, interfaces for the following parallel direct solvers are also implemented.

MUMPS is a parallel direct solver based in MPI implementations \cite{mumps_1,mumps_2}. The object-oriented approach of ParMooN has enabled the implementation of MUMPS in ParMooN without much overhead in computation and memory. 
 MUMPS is implemented in such a way that both the distributed and shared memory model from ParMooN can call. The system matrix and load vector is provided as an input to MUMPS in 
a distributed manner, $i.e.,$ each MPI processor maps its entries in the distributed system matrix to the global system matrix by the method discussed in section \ref{mappingMUMPS}. 

PastiX is a parallel direct solver, similar to MUMPS, based on MPI implementation. It was developed at inria labs \cite{pastix}.
While ParMooN exclusively uses a CSR data structure for storing matrices, PastiX requires the input to be provided in a CSC format. 
Considering the symmetry available in the FEM systems, the interface to the solver can be realized with little overhead. As discussed above, the global id of the DOFs help in providing the matrix input in a distributed format.

The computations are performed on the SahasraT XC40 machine at SERC, Indian Institute of Science, Bangalore. 
The SahasraT XC40 \cite{SahasraT} is an Intel Haswell $2.5$ GHz based CPU cluster with $1376$ nodes accounting to $33,024$ cores in total, and a memory of $128$ GB per node. For the comparison of solvers, the experiments are performed on the Tyrone cluster at SERC \cite{tyrone}.
This cluster is a heterogeneous cluster composed of two types of nodes, 9 nodes with 32-cores each and 8-nodes with 64-cores each. 
The 32-core node has a 2.4GHz AMD Opteron 6136 processor and 64GB RAM. The 64-core node has 2.2GHz AMD Opteron 6274 processor and 128GB RAM. 

Next, to quantify the parallel performance of the developed parallel scheme, the following parameters are calculated: \\
{\acrfont \textbf{Speedup}} : The ratio of the total time taken by the reference set of processors to the total time taken by a given set of processors. \\
{\acrfont \textbf{Ideal speedup}} : The ratio of the number of processors in a given set to the   number of processors in the reference set. \\
{\acrfont \textbf{Parallel efficiency}} : The ratio of the speedup to its ideal speedup.

\subsection{Model Problem}\label{TCD3D}
We consider the heat equation with the Dirichlet boundary condition
\begin{align*}
 \frac{\partial u}{\partial t}- \Delta u  &= f \quad \mbox{in } (0,T]\times\Omega,\\
  u(0,x,y,z) &= 0\ \ \ \mbox{in }\  \Omega,  
\end{align*}
as the model problem. Here, the used end time $T=5$ and   domain $\Omega:=(0,1)^3$. Further, the Dirichlet boundary value and the source term $f$  are chosen in such a way that the solution
\begin{align*}
  u(t, x,y,z) = e^{-0.1\,t}\sin(\pi x)\cos(\pi y)\cos(\pi z) 
\end{align*}
satisfies the heat equation. The domain is triangulated into tetrahedral cells. Further, the standard Galerkin finite element method and 
the Crank-Nicolson scheme are used for the spatial and temporal discretization, respectively. The used time step is 0.01 and it results in 500 time steps in total.
The Gauss--Seidel method is  used as smoother at all the levels of multigrid. Three smoothing iterations are performed on each 
  pre-  and post-smoothing steps. On the coarsest grid the Gauss--Seidel method is used to solve the system exactly.

\subsection{Comparison of ParMooN Multigrid and Direct Solvers}
The memory overhead in direct solvers are comparatively higher than the overhead in iterative solvers for a problem of same size. 
It is one of the major advantage of using iterative solvers over direct solvers, especially in large scale problems.  
 However, the iterative methods will be inefficient for solving system of   equations with multiple 
right hand sides (RHS). The direct solvers on the other hand factorize the system matrix only once, 
and the solution for multiple RHS  can be obtained by forward elimination and backward substitution. Direct solvers also 
prove to be efficient for time-dependent problems when the system matrix does not change in time as the system matrix needs to be factorized only once at the beginning. 

Two types of geometric grids are considered here. Grid~1 consists of  262,144 cells and  $P_1$ finite element is chosen on this grid. 
Grid~2 consists of 32,768 cells and $P_2$   finite element is chosen on Grid~2.
Both grids  contain 274625 DOFs.
\begin{figure}
\begin{center}
\includegraphics[trim={0cm 0 0 0},clip,width=0.36\textwidth,height=0.216\textwidth]{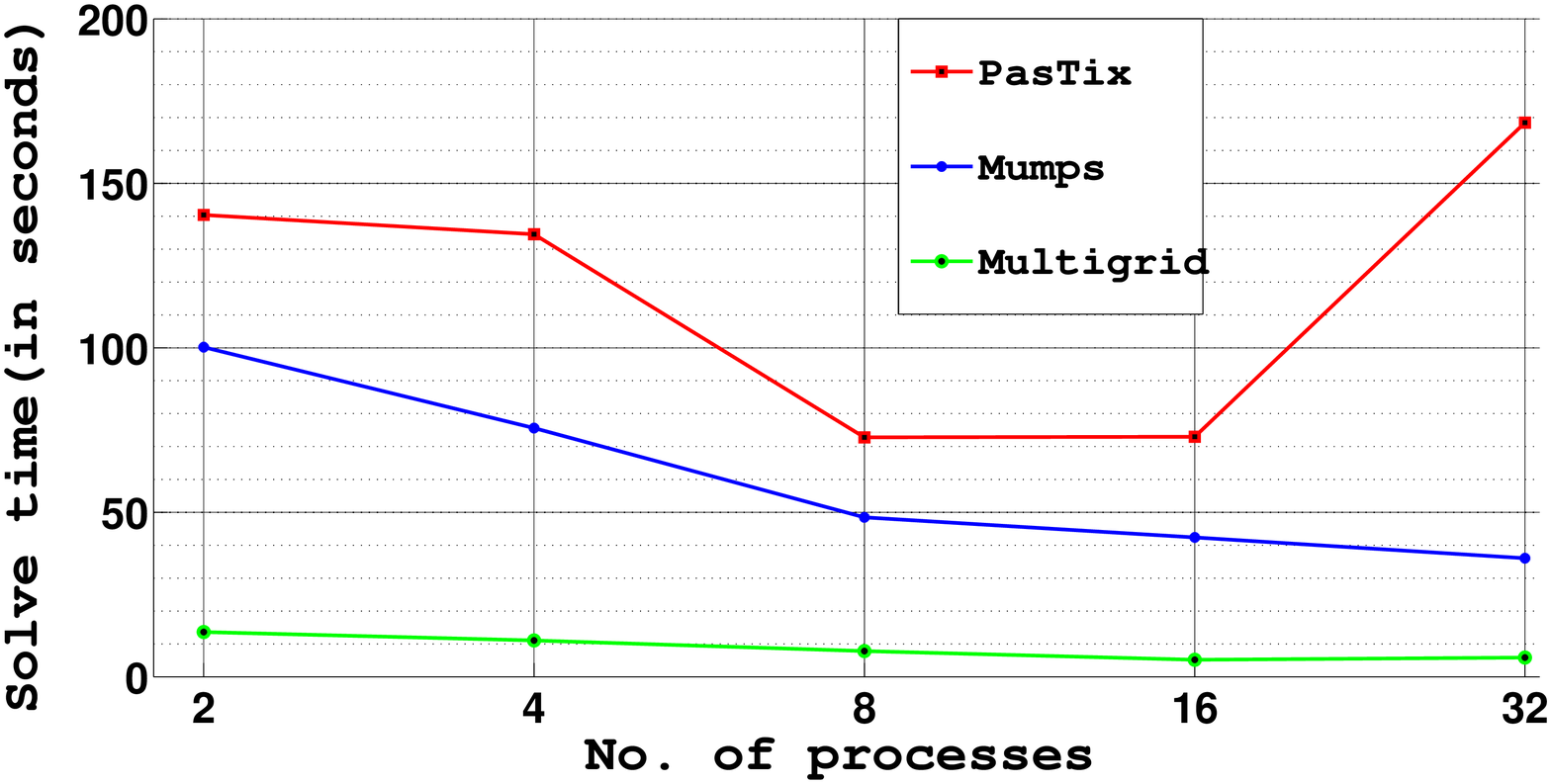}
\caption{Solve time with a P1 element on Grid1}
\label{fine_P1}
\end{center}
\end{figure}
\begin{figure}
\begin{center}
\includegraphics[trim={0cm 0 0 0},clip,width=0.36\textwidth,height=0.216\textwidth]{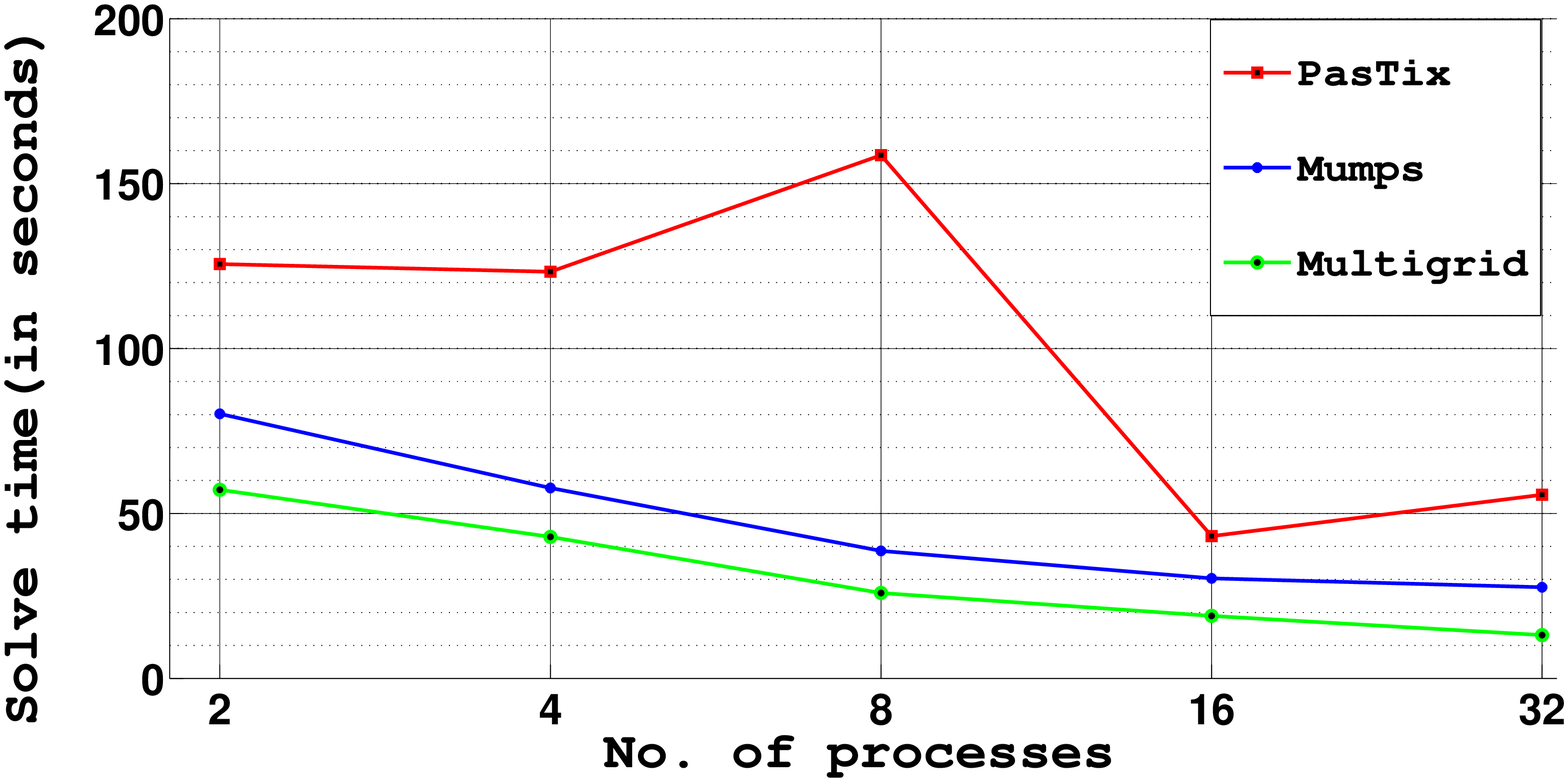}
\caption{Solve time with a P2 element on Grid2}
\label{coarse_P2}
\end{center}
\end{figure}
It can be seen from Figures \ref{fine_P1} and  \ref{coarse_P2}, the multigrid solver performs better among the considered direct solvers.
 Also observed is that the compute time involved is lower when higher order elements (P2 element) are considered, with the same system size.
 This is because of a better rate of convergence (fewer number of iterations) with higher order finite elements. However, it has an adverse effect on the direct solvers as higher order elements relatively decrease the sparsity of the system. This could result in increased fill-in  during the factorization step and thereby higher compute time with direct methods.

\subsection{Performance of ParMooN}
\begin{figure}
\begin{center}
\includegraphics[width=0.5\textwidth]{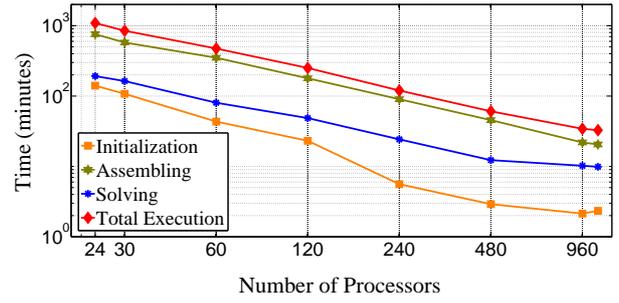}
\caption{Time taken for initialization, assembling and solving the model problem with $135,005,697$ DOFs.}
\label{Heat_Executuion_Time}
\end{center}
\end{figure}

\begin{figure}
\begin{center}
\includegraphics[width=0.5\textwidth]{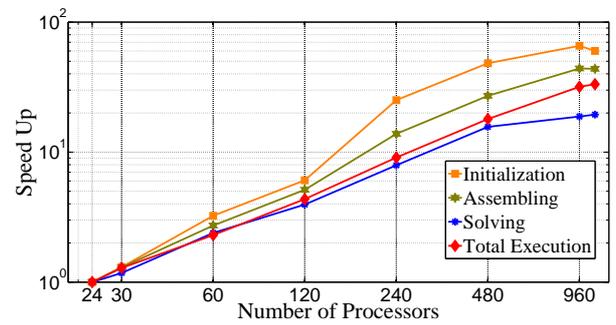}
\caption{Speedup obtained  in  model problem with $135,005,697$ DOFs.}
\label{Heat_SpeedUp}
\end{center}
\end{figure}

\begin{figure}
\begin{center}
\includegraphics[width=0.5\textwidth]{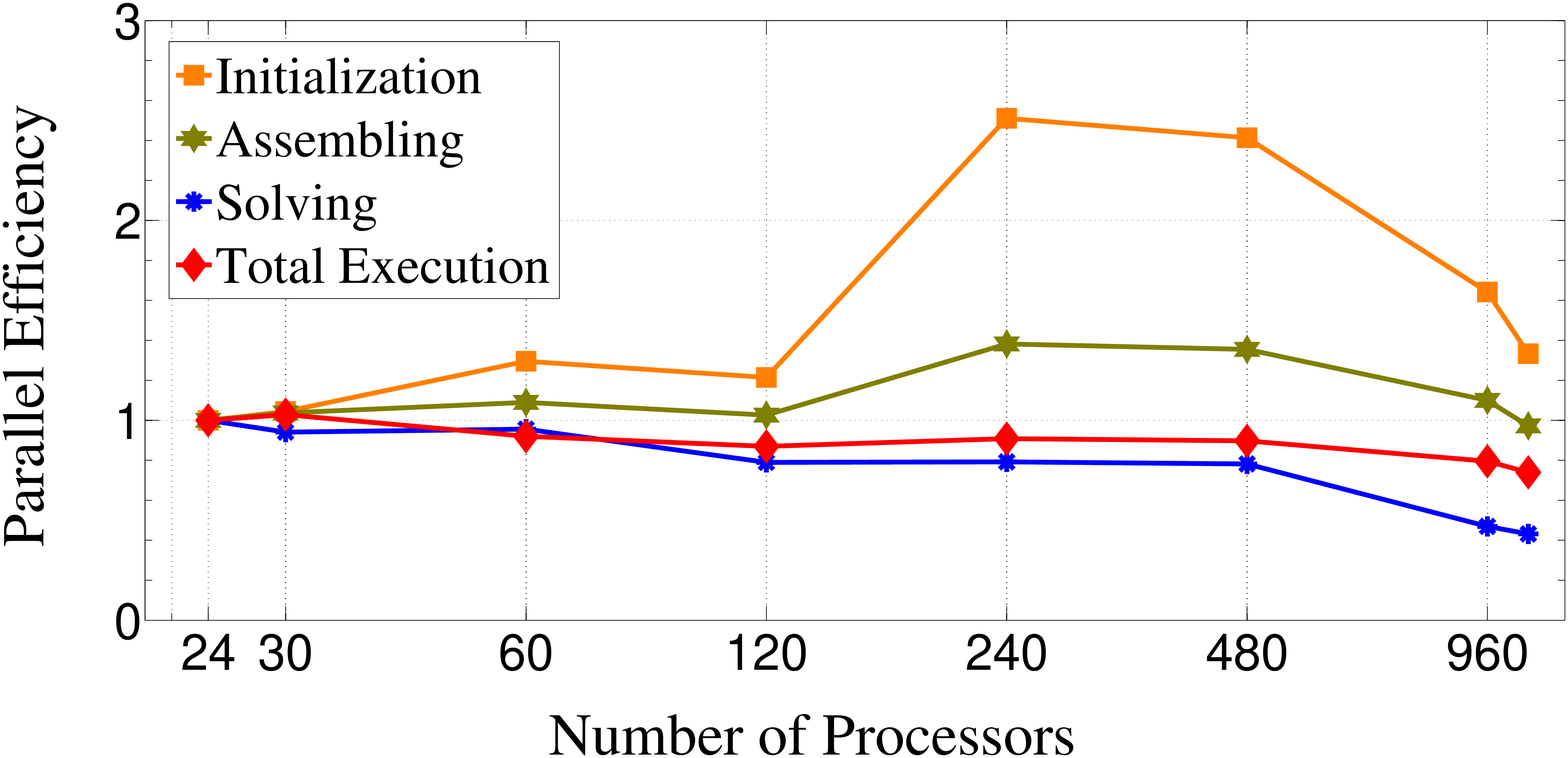}
\caption{Parallel Efficiency for the model problem  with $135,005,697$ DOFs.}
\label{Heat_Parallel_Efficiency}
\end{center}
\end{figure}

We finally perform an array of computations for  the model problem with different number of MPI processors on the SahasraT machine. In this study,
a hierarchy of six multigrid levels with piecewise linear finite element, $P_1$, and up to 1080 MPI processors are used. The finest level consists of $805,306,368$ 
cells and $135,005,697$ DOFs, whereas the coarsest level consists of $24,576$ cells and $4,913$ DOFs. 
Figures \ref{Heat_Executuion_Time}, \ref{Heat_SpeedUp} and \ref{Heat_Parallel_Efficiency} show the time taken, speedup and parallel efficiency for initialization, 
assembling the system matrix, solving the system  and the total execution time by different number of MPI processors.  The initialization step consists  of allocation
of memory, construction of ParFEMapper and ParFECommunicator. The time taken for assembling the mass, stiffness matrices and load vector is the assembling time in Fig.~\ref{Heat_Executuion_Time}.  The time spent by the multigrid solver including the communication time during restriction and prolongation steps is termed as a solving time.  The  total execution time is the total time taken for solving the entire problem. The total execution time is reduced from $18.17$ hours with $24$ processors to $0.55$ hours with $1080$ processors. The algorithm scales up very well to $960$ processors. 

The initialization step has two expensive steps, the master-slave DOF verification step and the mapping step. These steps depend on the inter-process communication, 
and the sending and receiving  message size for mapping between neighboring processors will be huge when fewer processors are used.
With a huge increase in the number of processors, the sending and receiving  message size between neighboring processors decreases, and a much faster communication is observed. 
Further, a large sized memory allocation is performed while using fewer processors, and thus increasing the initialization cost. 
However,  the parallel efficiency of the initialization decreases when the number of MPI processors is kept on increasing  as   the very smaller size message becomes communication intensive.
Nevertheless, the parallel efficiency of the initialization step is more than one even for $1080$ processors. Note that the initialization step is a one step process,
and still is very efficient. Hence, it is not a major concern for further scaling of the proposed algorithm. 

Next, the assembling step is parallel efficient as expected, since assembling does not require any communication. Moreover,  
the super linear curve of assembling can be  attributed to the cache effect. The challenging step with respect to the scaling 
of the algorithm is the solver. Multigrid is very efficient compared to MUMPS as seen earlier in Figure \ref{coarse_P2} 
but fails to scale as similarly as initialization and assembling.  Even with communication, the time taken by the solver is much less than the assembling time.
It gives an indication that the considered problem is not computationally intensive, as assembling has $\mathcal{O}(n)$ complexity. Nevertheless, the scaling of solving time is good up to 960 processors.
Scalar problems are not so computationally expensive compared to  Navier--Stokes problems in 
higher dimensions. The algorithm is expected to show better scaling when solving vector problems like Navier--Stokes models. 
Table~\ref{HeatEquation_Execution_Time_Table} 
shows that for $1080$ processors more time is spent on 
communicating rather than solving, and thus the solver is not expected to scale any further. It is due to the fact that the entire mesh on the coarsest level
has only $4,913$ DOFs, and as a result  only $4$ to $5$ own DOFs on each MPI processor  while using $1080$ processors. 
The multigrid method spends most of its time on the coarser levels and the algorithm becomes communication intensive and computationally
less intensive as we move towards coarser levels. Hence, the algorithm is expected to suffer while using higher number of processors if
the coarser levels do not possess sufficiently many DOFs.

 \begin{table*}[t]
\centering
\begin{tabular}{|p{1cm}|p{1.8cm}|p{1.7cm}|p{1.2cm}|p{2.3cm}|p{1.cm}|p{1.2cm}|p{2.5cm}|} \hline
 Total MPI Ranks  & \hspace{.8in} Initialization &\hspace{.8in} Assembling \centering (A) &\hspace{.8in} Solving \centering (S) & \hspace{.8in}Communication \centering (C) & \hspace{.35in} S\&C &\hspace{.35in} A,S\&C & \hspace{.9in}Total Execution \\ \hline
24&140.48&757.41&186.25&5.82&192.08&949.49&1089.97 \\ \hline
30&107.62&576.76&158.38&4.95&163.32&740.08&847.70 \\ \hline
60&43.37&350.02&78.05&2.33&80.38&430.40&473.77 \\ \hline
120&23.13&178.86&46.30&3&2.33&227.50&250.63 \\ \hline
240&5.59&90.17&22.12&2.14&24.24&114.41&120.00 \\ \hline	
480&2.91&45.51&11.15&1.13&12.28&57.79&60.70 \\ \hline
960&2.14&21.91&6.72&3.51&10.23&32.14&34.27 \\ \hline
1080&2.34&20.52&5.53&4.38&9.88&30.40&32.74 \\ \hline
\end{tabular}
\caption{Execution time in seconds for the model problem.}
\label{HeatEquation_Execution_Time_Table}
\end{table*}
 
\section{Summary}
 Objected-oriented parallel finite element algorithms with a data structure to handle geometric multigrid method have been proposed. The proposed parallel implementation supports 
 hybrid MPI-OpenMP computations. The design and implementation of two classes, ParFEMapper and ParFECommunicator that handle the mapping and communication routines across all MPI processors are discussed in detail. 
 The proposed parallel finite element solver was compared  with the  parallel direct solvers MUMPS and PasTiX. 
 The performance of the solver was analyzed and a good speedup was observed for a reasonable problem size. 
 More performance analysis for computationally intensive models such as Navier-Stokes problems will be  part of our future work.

\section*{Acknowledgment}
The authors would like to thank Supercomputer Education and Research Centre, Indian Institute of Science (IISc), Bangalore for proving  access to the supercomputer SahasraT. 

 \bibliographystyle{elsarticle-num}
\bibliography{main}

%
%

\end{document}